\documentclass[
a4paper,
preprint,
aps,
prl,
showpacs,
superscriptaddress,
longbibliography
]{revtex4-1}

\usepackage{natbib}
\usepackage{graphicx}
\usepackage{bm}
\usepackage{amsmath,amssymb}
\usepackage{eucal}
\usepackage{soul}
\usepackage{color}

\begin{document}
	
	\title{Widom delta of supercritical gas-liquid coexistence}
	
	\date{\today}
	
	\author{Min Young Ha}
	\affiliation{School of Chemical and Biological Engineering, Institute of Chemical Processes, Seoul National University, Seoul 08826, Republic of Korea}
	\author{Tae Jun Yoon}
	\affiliation{School of Chemical and Biological Engineering, Institute of Chemical Processes, Seoul National University, Seoul 08826, Republic of Korea}
	\author{Tsvi Tlusty}
	\email{tsvitlusty@gmail.com}
	\affiliation{Center for Soft and Living Matter, Institute for Basic Science (IBS), Ulsan 44919, Korea}
	\affiliation{Department of 	Physics, Ulsan National Institute of Science and Technology (UNIST), Ulsan 44919, Korea}
	\author{Yongseok Jho}
	\email{ysjho@gnu.ac.kr}
	\affiliation{Department of Physics and Research Institute of Natural Science, Gyeongsang National University, Jinju 52828, Republic of Korea}
	\author{Won Bo Lee}
	\email{wblee@snu.ac.kr}
	\affiliation{School of Chemical and Biological Engineering, Institute of Chemical Processes, Seoul National University, Seoul 08826, Republic of Korea}
	\date{\today}
	
	\begin{abstract}
		We report on the coexistence of liquid-like and gas-like structures in supercritical fluid (SCF). The deltoid coexistence region encloses the Widom line, and may therefore be termed the ``Widom delta". Machine learning analysis of simulation data shows continuous transition across the delta, from liquid-like to gas-like states, with fractions following a simplified two-state model. This suggests a microscopic view of the SCF as a mixture of liquid-like and gas-like structures, where the anomalous behavior near the critical point originates from fluctuations between the two types.
		
	\end{abstract}
	
	
	\maketitle
	\textit{Introduction.--}  
	Ever since its discovery, the liquid-gas critical point (LGCP) has been central to the progress of thermodynamics and statistical physics \cite{stanley1971phase}. 
	At the LGCP, the boiling curve -- separating the liquid and gas phases -- ends, and the two phases merge into a single phase of supercritical fluid (SCF). Understanding the SCF has also practical implications, since the blend of liquid-like and gas-like traits renders it particularly useful in chemical engineering  ~\cite{Eckert1996SupercriticalProcessing,Reverchon1997SupercriticalProducts,Savage1995ReactionsFundamentals,Niessen2007}. 
	
	In the pressure-temperature phase diagram, a bypass around the LGCP via the SCF will continuously transform liquid into gas, thereby avoiding a first-order boiling transition. The SCF was therefore regarded as a homogeneous state of matter. But recently, a series of seminal works have accumulated mounting evidence of non-monotonic behavior and detectable crossovers in salient features of the SCF ~\cite{Xu2005RelationTransition,McMillan2010FluidSupercritical,Simeoni2010TheSupercriticalfluids,Brazhkin2011WidomSystem,Gallo2014WidomWater.}. 
	All this suggested the existence of a boundary separating the SCF into gas-like and liquid-like regions.
	
	A prominent candidate boundary is the \textit{Widom line}, introduced by Stanley \textit{et al.} as the locus of maximum correlation length ~\cite{Xu2005RelationTransition,McMillan2010FluidSupercritical}. At the Widom line, thermodynamic response functions show local maxima ~\cite{ Brazhkin2011WidomSystem,Xu2005RelationTransition}, and the dispersion of sound undergoes a sharp transition~\cite{Simeoni2010TheSupercriticalfluids}. 
	These hallmarks are reminiscent of the divergent discontinuity at the vapor-liquid equilibrium (VLE) line. Hence, the Widom line is often regarded as an extension of the VLE line into the SCF. In this viewpoint, the Widom line divides the SCF into two distinct bulk states, which are entirely liquid-like or gas-like, though the singularity at the transition is weaker than that of first and second-order phase transitions. The questions regarding the physical nature of the boundary and of the states it defines -- in particular their structural features -- remain open. 
	
	This Letter addresses these questions and proposes an alternative picture: In contrast to the view that liquid and gas can only be mingled during the non-equilibrium boiling process along the VLE line~\cite{Jeong2015Micro-structuralMorphology}, we demonstrate the coexistence of two local states of molecules, liquid-like and gas-like, within a wide region of the SCF enclosing the Widom line. The SCF appears as an inhomogeneous mixture of liquid-like and gas-like particles, and the continuous transition from SCF to pure liquid or gas can be understood as the gradual vanishing of this liquid-gas coexistence.
	
\begin{figure*}
		\centering
		\includegraphics[width=17.0cm]{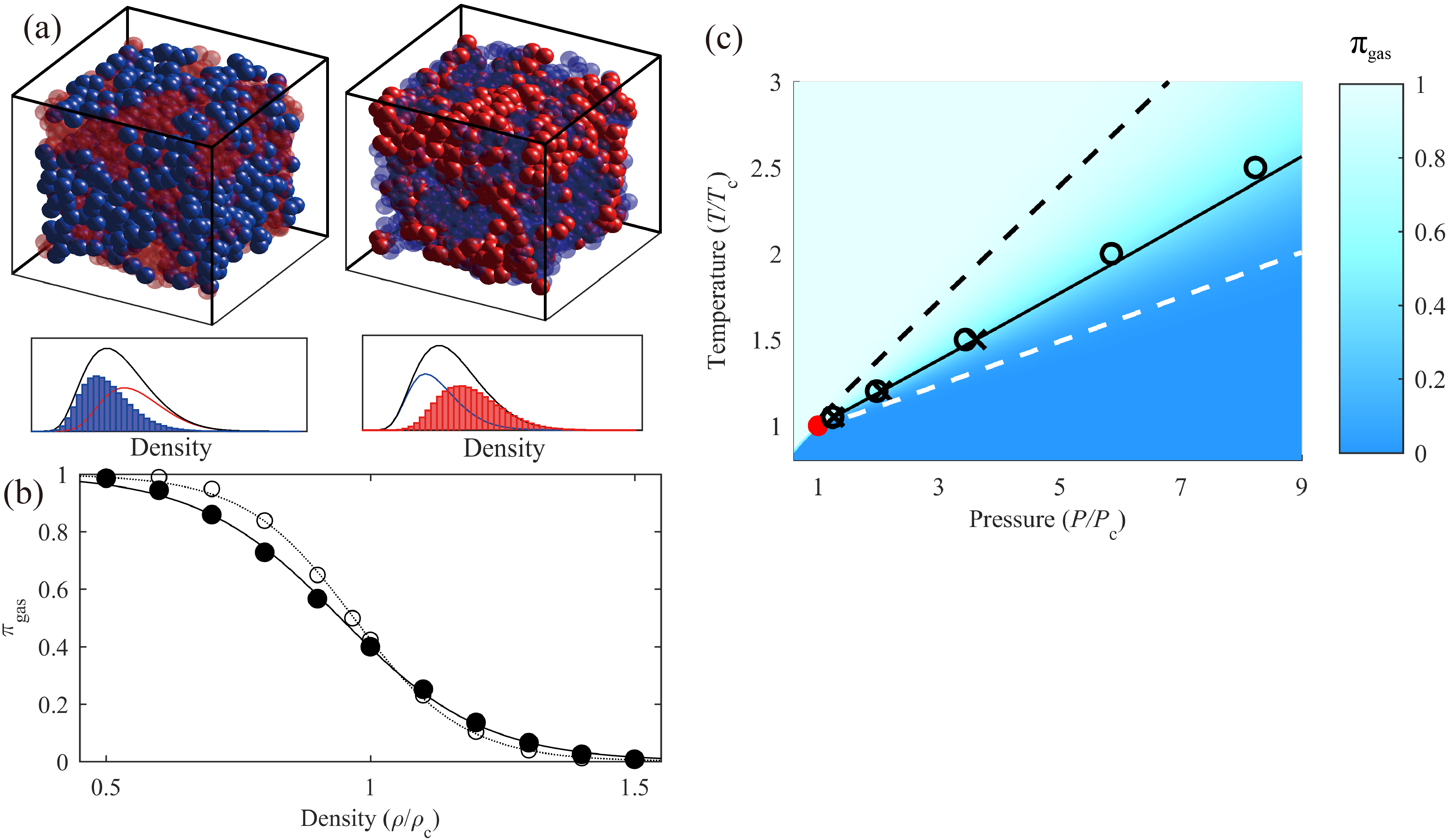}
		\caption{ \textbf{The Widom delta.} 
			(a) A snapshot of the Lennard-Jones supercritical fluid (SCF),
			at $T=1.05T_c$ and $P=1.24P_c$, shows an inhomogeneous mixture of gas-like (blue spheres) and liquid-like (red spheres) particles, as classified by the neural network (Methods). These two species  form loosely connected, intertwined clusters. For clarity, either gas-like (right) or liquid-like (left) particles are semi-transparent. The histograms depict the local density distribution of gas-like (blue) and liquid-like (red) particles in the SCF, acquired from the inverse volume of the Voronoi cells.
			(b) The proportion of gas-like particles, $\pi_{\textrm{gas}}$, for the $T=1.05T_c$ (filled circles) and $T=9.0T_c$ (empty circles) isotherms. The solid and dotted lines are fits to the sigmoidal curve, Eq.~(\ref{eq:sigmoid}).
			(c) The Widom delta in the pressure-temperature phase diagram, which shows points where $\pi_{\textrm{gas}}=0.5$ (circles), local maxima of the specific heat, $C_P$ (crosses), the critical isochore (solid line),  the loci of $\pi_{\textrm{gas}}=0.95 \textrm{ and } 0.05$ (dashed lines), and the critical point (red circle).
		}
		\label{fig:Abstract}
	\end{figure*}

	The deltoid coexistence region emanates from the critical point, akin to a river delta emanating at the river mouth. One may therefore call it the ``\textit{Widom delta}", and a representative example is shown in Fig.~\ref{fig:Abstract}(c). We propose that the anomalous behavior of the SCF at the tip of the delta, close the LGCP, has its roots in the fluctuation between liquid-like and gas-like microstates.

	
	\textit{Methods: Machine learning.--}
	Since liquid-like and gas-like particles are mixed in a single phase of SCF, the local states cannot be classified by standard order parameters, such as the density in liquid-vapor phase transition. To overcome this obstacle, we adopted a machine learning approach using deep neural networks (DNN), which allows us to capture subtle structural features by combining multiple parameters of local arrangement ~\cite{Behler2007GeneralizedSurfaces,Cubuk2015IdentifyingMethods,Schoenholz2015ALiquids,Carrasquilla2016MachineMatter,vanNieuwenburg2017LearningConfusion}. The DNN was trained with local structure data of near-critical liquid and gas particles, acquired from molecular dynamics (MD) simulation of Lennard-Jones (LJ) fluid at $T=0.97T_c$. The training and validation accuracies were ${\sim}95\%$, implying effective training and generalization (see further details in the Supplementary Material).

	\textit{Results: the Widom delta.--}
	To map and characterize the coexistence region, we employed the trained DNN to label individual particles in supercritical LJ fluid as liquid-like or gas-like. A representative snapshot is shown in Fig.~\ref{fig:Abstract}(a), where the two types of particles form distinct domains: microstates of intertwined clusters, which are not phase-separated nor uniformly mixed. 
	Keeping in mind that no collective input was used to train the DNN, but only the labels of individual particles, the patterns observed in the simulation hint for an underlying physical origin of the particle-based classification.
	Moreover, these structural patterns suggest that liquid-like clusters are the origin of local density augmentation, a hallmark of the SCF \cite{lewis2001local}.
	
	We further examined the supercritical coexistence in terms of a simplified two-state model: The SCF is partitioned into fractions $\pi_{\textrm{gas}}$ of gas-like particles and $\pi_{\textrm{liq}}=1-\pi_{\textrm{gas}}$ of liquid-like particles. The overall free energy of the SCF, $G$ , augments the energy with the entropy of mixing, 
	\begin{equation}
	G = \pi_{\textrm{liq}}  G_{\textrm{liq}} + \pi_{\textrm{gas}} G_{\textrm{gas}}
	+ k_{\rm B}T\left[\pi_{\textrm{liq}}  \ln{ \pi_{\textrm{liq}}}+\pi_{\textrm{gas}} \ln {\pi_{\textrm{gas}} }\right]~,
	\end{equation}
	where $G_{\textrm{liq}}$ and $G_{\textrm{gas}}$ are the energies of the two states. 
	In equilibrium, minimization of $G$ yields the familiar Fermi-Dirac distribution,
	\begin{equation}
	\pi_{\textrm{gas}} = \left[1+e^{-\Delta G} \right]^{-1},~
	\pi_{\textrm{liq}} = \left[1+e^{\Delta G} \right]^{-1},
	\end{equation}
	where $\Delta G = (G_{\textrm{gas}}-G_{\textrm{liq}})/(k_{\rm  B} T)$. Near the critical point, we may expand $\Delta G$ around the critical pressure and density  $\Delta G \sim P-P_c  \sim \rho- \rho_{c}$ (at constant temperature), and obtain a sigmoidal curve of the fraction $\pi_{\textrm{gas}}$ as a function of the bulk density, $\rho$,
	\begin{equation}
	\pi_{\textrm{gas}} = \left[ 1 + a e^{b\rho} \right]^{-1},
	\label{eq:sigmoid}
	\end{equation}
	where $a$ and $b$ are fit parameters (Fig.~\ref{fig:Abstract}(b)).
	Despite the simplicity of the coarse-grained two-state model, it exhibits excellent fit ($R^2>0.99$) up  to a temperature  $T=9.0T_c$.  This demonstrates a well-defined dependence of $\pi_{\textrm{gas}}$ and $\pi_{\textrm{liq}}$ on the density (or the pressure). 
	
	We note that we previously applied a statistical mixture model to examine the possibility of supercritical coexistence in  $\textrm{CO}_2$~\cite{Yoon2017MonteTessellation,Yoon2017MolecularNaphthalene}. However, we could not directly observe any inhomogeneity because we were unable to assign individual molecules to microstates. 
	In contrast, the present machine-learning approach can label every molecule as liquid-like or gas-like, and both thermodynamic and kinetic characteristics of the microstates can be quantified and compared:
	
	\begin{figure}
		\centering
		\includegraphics[width=8.6cm]{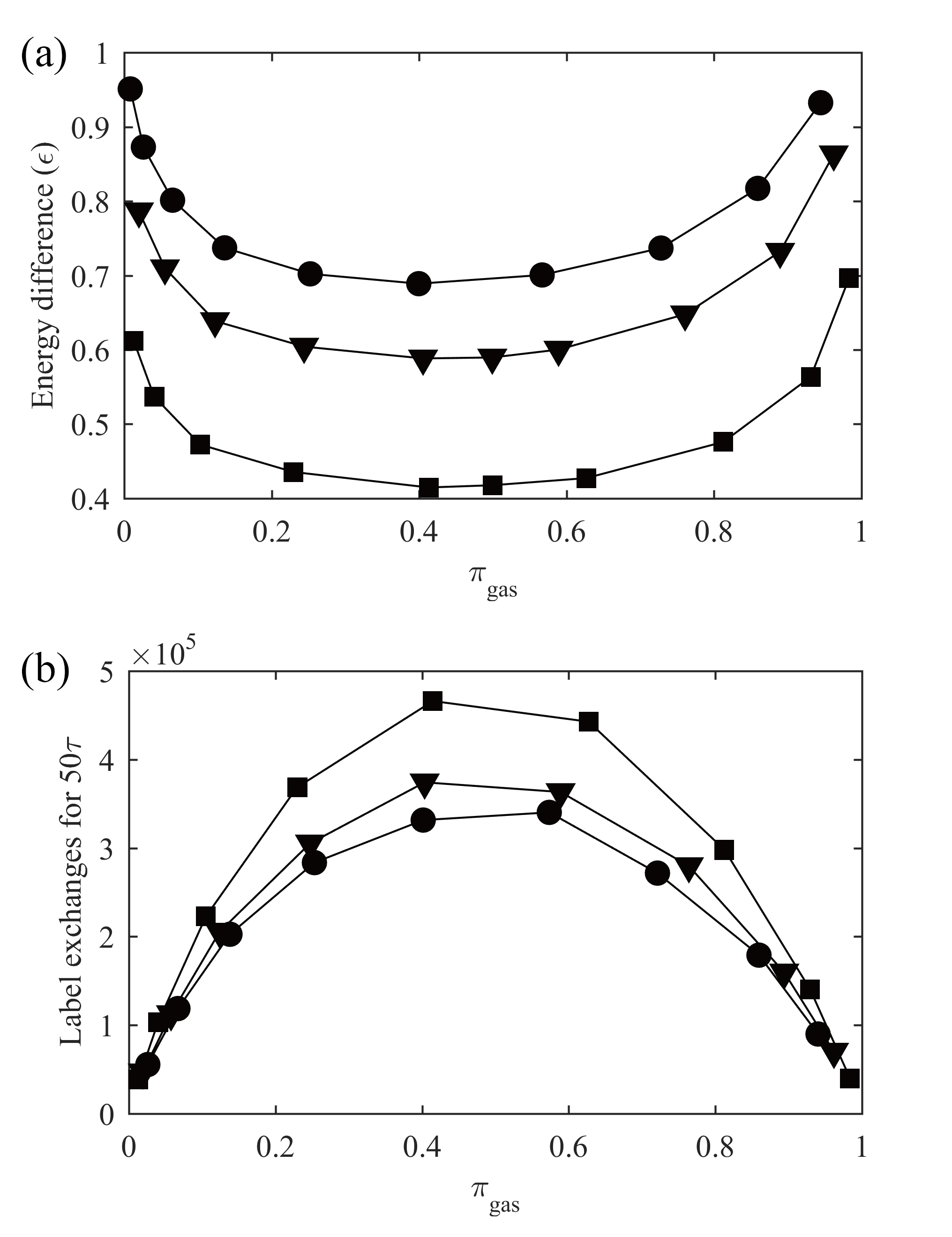}
		\caption{\textbf{Energetics and kinetics in the supercritical coexistence region.}
			(a) Difference in potential energy, between liquid-like and gas-like particles along the isotherms  $T=1.05T_c$ (circles),  $T=1.2T_c$ (triangles), and $T=2.0T_c$ (squares). The units are expressed in the Lennard-Jones parameter $\epsilon$. The solid lines are guide for the eye.
			(b) Counts of label exchange events for 50$\tau$ for the same isotherms, for $\tau=(m\sigma^2/\epsilon)^{1/2}$ for LJ parameters $\epsilon$, $\sigma$ and particle mass $m$.
		}
		\label{fig:Cluster}
	\end{figure}

	Isotherms of the potential energy difference between liquid-like and gas-like particles exhibit minima  near $\pi_{\textrm{gas}}=\pi_{\textrm{liq}}=0.5$ (Fig.~\ref{fig:Cluster}(a)). The lower energy barrier induces higher rates of exchange between liquid-like and gas-like particles,  which are indeed most frequent in this region (Fig.~\ref{fig:Cluster}(b)). The fluctuations between the supercritical microstates are most vigorous at the loci of $\pi_{\textrm{gas}}=0.5$, and decay towards the pure states, $\pi_{\textrm{gas}}=0$ (liquid) or $\pi_{\textrm{gas}}=1$ (gas). Owing to the low energy barrier at the $\pi_{\textrm{gas}}=0.5$ point, the fluid becomes most susceptible to perturbations.
	
	\begin{figure}
		\centering
		\includegraphics[width=8.6cm]{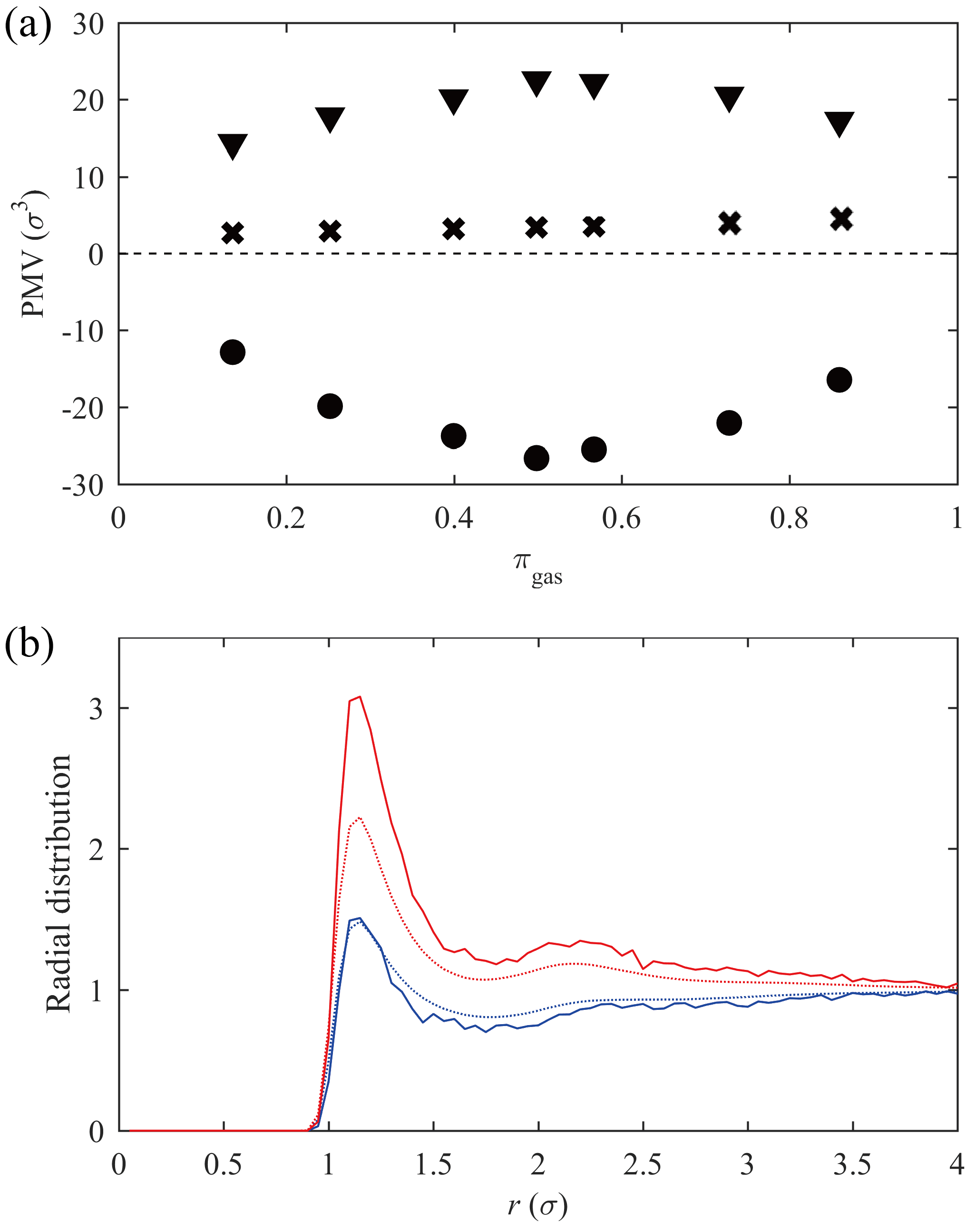}
		\caption{\textbf{Anomalous behavior of the supercritical fluid.}
			(a) Partial molar volume (PMV) of attractive solute (circles), repulsive solute (triangles) and solvent particles (crosses) in SCF at $T=1.05T_c$. The attractive and repulsive solutes were modeled as Lennard-Jones particles, with LJ parameters $\epsilon_\textrm{att}=2.0\epsilon_{\textrm{solvent}}$, $\epsilon_\textrm{rep}=0.5\epsilon_{\textrm{solvent}}$, and $\sigma_\textrm{att}=\sigma_\textrm{rep}=\sigma_{\textrm{solvent}}$. Lorentz-Berthelot mixing rule was used to account for interactions between different LJ particle pairs.
			(b) Radial distribution functions (RDFs) of liquid-like (red) and gas-like solvent particles (blue) around solvent particles (dotted lines), and around attractive solute particles (solid lines).
		}
		\label{fig:RDF}
	\end{figure}
	
	Next, we test the possible link between the coexistence and anomalous behavior of the SCF. 
	One such hallmark of supercritical solutions is the divergence of the solute's partial molar volume (PMV) close to the critical point ~\cite{Eckert1986SoluteFluids,Petsche1989Solute-solventInvestigation,Zhang2002StudyCompressibility,Zhang2002Effect,Stubbs2005PartialStudy}. 
	In particular,  the PMV of attractive solutes is known to exhibit largely negative PMV, \textit{i.e.} the total volume of the solution substantially decreases upon the addition of solute.
	Within the supercritical coexistence scenario, one may interpret this phenomenon as the transition between microstates induced by an external perturbation, namely, the addition of solutes. Thus, we expect to observe extrema of PMV around $\pi_{\textrm{gas}}=0.5$. Fig.~\ref{fig:RDF}(a) is consistent with this prediction, and shows that the PMV magnitude of both the attractive and repulsive solutes are indeed maximal at $\pi_{\textrm{gas}}=0.5$.
	
	The anomalous behavior shows also a structural manifestation in the radial distribution functions (RDF) of liquid-like and gas-like solvent particles around solvent and solute particles (Fig.~\ref{fig:RDF}(b)). The concentration of liquid-like particles was greatly enriched around an attractive solute. This could be the result of an effective pressure exerted by the solute, which may induce the transformation of nearby gas-like solvent particles into liquid-like particles. This mechanism induces negative PMV and increases the density.
	
	To conclude, it appears that the point of $\pi_{\textrm{gas}}=0.5$ is where (i) the fluctuation between microstates is most vigorous, and as a result (ii) the fluid is most susceptible. The latter implies that the thermodynamic response functions should have their local maxima at $\pi_{\textrm{gas}} \simeq 0.5$. This leads us to suggest an alternative, practical definition of the the Widom line as the locus of $\pi_{\textrm{gas}}=0.5$. Fig. \ref{fig:Abstract}(c) depicts the phase diagram of LJ fluid, where the $C_P$-Widom line (maximal isobaric specific heat), the critical isochore, and the line of $\pi_{\textrm{gas}}=0.5$ are shown. It was previously demonstrated that the $C_P$-Widom line of LJ fluid follows its critical isochore~\cite{Brazhkin2011WidomSystem,Artemenko2017TheFluids}, and the newly defined Widom line lies close to both. However, unlike the $C_P$-Widom line which is rapidly smeared out at high temperature, the $\pi_{\textrm{gas}}=0.5$ line remains observable  even in the deeply supercritical region, which allows us to draw a global line in the phase diagram.
	
	The SCF-gas and SCF-liquid boundaries can be marked by the (effective) disappearance of the liquid-gas coexistence. In Fig.~\ref{fig:Abstract}(c), these borders are marked by the lines of $\pi_{\textrm{gas}}=0.05$ (SCF-liquid) and  $0.95$ (SCF-gas). We emphasize, however, that these values were chosen since the training accuracy was approximately $~95\%$, and classification beyond these lines is not robust enough. We therefore cannot exclude the possibility that the supercritical coexistence region, \textit{i.e.} the Widom delta, further extends.  
	Within the supercritical coexistence model, liquid and gas can be continuously transformed into each other  by detouring the critical point. Along this pathway, the system crosses two mild boundaries without latent heat, where the smooth change of $\pi_{\textrm{gas}}$ and $\pi_{\textrm{liq}}$ allows one to avoid the abrupt first-order transition. 
	
	\textit{Discussion.--}
	In the present Letter, the application of machine learning techniques demonstrates that supercritical fluid (SCF) can be seen as an inhomogeneous mixture of liquid-like and gas-like particles. Below the critical point, the liquid-gas mixture would be unstable due to large surface energy, and the vapor-liquid equilibrium (VLE) line completely separates the two phases after equilibration. But as the temperature exceeds the critical point, the VLE line broadens into a two-dimensional region (the ``Widom delta") in which the SCF emerges as an inhomogeneous mixture of coexisting liquid-like and gas-like molecules.  In the Widom delta, instead of undergoing macroscopic phase separation, liquid-like and gas-like particles form microscopically intertwined domains. Hence, the emergent SCF shows mild and continuous responses to temperature and pressure variations, without latent heat or jump discontinuities.
	
	The Widom line can be reinterpreted on a microscopic basis, as a collection of points where the proportion of liquid-like and gas-like molecules is even. Moreover, the Widom line is where the transitions between microstates are strongly enhanced, and the fluid is therefore maximally susceptible to external perturbations. This may be the origin of the bulk crossover phenomena and the local maxima of the response functions, which are reminiscent of the subcritical VLE line.
	
	Finally, the proposed microscopic understanding of the SCF may provide insights into critical phenomena, and trigger further theoretical developments. One possible future direction is the examination of liquid-liquid criticality. The machine learning scheme can be used to capture the high-density and low-density domains in supercooled liquid, which may advance the understanding of exotic properties of ambient water.
	
	\begin{acknowledgments}
		This work was supported by National Research Foundation of Korea, grants NRF-2015R1A2A2A01007379 and NRF-2017H1A2A1044355 (Global Ph.D. Fellowship Program), and by the Institute for Basic Science, grant IBS-R020.
	\end{acknowledgments}

\end{document}